\documentclass[universe,review,accept,moreauthors,pdftex]{Definitions/mdpi} 
\usepackage{aas_macros}
\newcommand{\gr}{$\gamma$-ray}
\newcommand{\grs}{$\gamma$-rays}
\newcommand{\nup}{$\nu_{\rm peak}^S$}

\newcommand{\lsim}{{\lower.5ex\hbox{$\; \buildrel < \over \sim \;$}}}
\newcommand{\gsim}{{\lower.5ex\hbox{$\; \buildrel > \over \sim \;$}}}
\newcommand{\nufnu}{$\nu$f$({\nu})$}

\firstpage{1} 
\pubvolume{1}
\issuenum{1}
\articlenumber{0}
\pubyear{2021}
\copyrightyear{2021}
\externaleditor{Academic Editor: Fridolin Weber 
} 
\datereceived{12 November 2021} 
\dateaccepted{9 December 2021} 
\datepublished{} 
\hreflink{https://doi.org/} 

\Title{Astrophysical Neutrinos and Blazars}

\TitleCitation{Astrophysical Neutrinos and Blazars}

\Author{Paolo Giommi 
 $^{1,2,3,}$*$^{,\dagger}$ 
\orcidA{} and Paolo Padovani$^{4,5,}$*$^{,\dagger}$\orcidB{}{}}

\AuthorNames{Paolo Giommi and Paolo Padovani}

\AuthorCitation{Giommi, P.; Padovani, P.}

\address{$^{1}$ \quad Institute for Advanced Study, Technische Universit{\"a}t M{\"u}nchen, Lichtenbergstrasse 2a, \mbox{D-85748 Garching bei M\"unchen
, Germany} \\
$^{2}$ \quad Center for Astro, Particle and Planetary Physics (CAP3), New York University Abu Dhabi, \mbox{Abu Dhabi P.O. Box 129188}, United Arab Emirates \\
$^{3}$ \quad Associated to Italian Space Agency, ASI, via del Politecnico snc, I-00133 Roma, Italy\\
$^{4}$ \quad European Southern Observatory, Karl-Schwarzschild-Str. 
2, D-85748 Garching bei M\"unchen, Germany\\
$^{5}$ \quad Associated to INAF-
Osservatorio di Astrofisica e Scienza dello Spazio, Via Piero 
Gobetti 93/3, \mbox{I-40129 Bologna}, Italy\\}
\corres{Correspondence: giommipaolo@gmail.com (P.G.); ppadovan@eso.org (P.P.)}

\firstnote{These authors contributed equally to this work.}

\abstract{We review and discuss recent results on the search for correlations between astrophysical neutrinos and \gr-detected sources, with many extragalactic studies reporting potential associations with different types of blazars. We investigate possible dependencies on blazar sub-classes by using the largest catalogues and all the multi-frequency data available. Through the study of similarities and differences in these sources we conclude that blazars come in two distinct flavours: LBLs and IHBLs (low-energy-peaked
and intermediate-high-energy-peaked objects). These are distinguished by widely different properties such as the overall spectral energy distribution shape, jet speed, cosmological evolution, broad-band spectral variability, and optical polarisation properties.
Although blazars of all types have been proposed as neutrino sources, evidence is accumulating in favour of IHBLs being the counterparts of astrophysical neutrinos. If this is indeed the case, 
we argue that the peculiar observational properties of IHBLs may be indirectly related to proton 
acceleration to very high energies.}


\begin{document}
\section{Introduction}

The discovery of ultra-high energy cosmic rays (UHECRs) established the existence of powerful cosmic accelerators capable of reaching energies millions of times larger than those that can be achieved by the best accelerators on Earth (see Ref. \citep{2019PhR...801....1A} and references therein).
The interaction of very high-energy cosmic rays (CRs) with matter or radiation in astrophysical contexts results in the production of neutral and charged mesons, which then decay into \grs, high-energy neutrinos and energetic particles, which lose energy in a variety of ways. Neutrinos, \grs\, (of energy \lsim 1 TeV, if not absorbed inside the emitting region or in the host galaxy), and the electromagnetic radiation emitted by secondary particles, are the `messengers'  that can reach the Earth undeflected from cosmological distances, providing information about the Universe at extreme energies.

A flow of high-energy neutrinos likely generated in these environments has been detected by the IceCube Observatory 
 (\url{https://icecube.wisc.edu}, accessed on 11 December 2021) 
  at the South Pole with energies extending beyond 1 PeV and an energy flux comparable to that observed in the \gr-band \citep{IceCube2013,2019ICRC...36.1004S,2020PhRvL.124e1103A}. This facility, however, cannot provide an accurate estimate of the arrival directions of the incoming neutrinos and is not sensitive enough to ensure firm detections of point sources with multiple events. Nevertheless, the absence of any significant anisotropy in the arrival direction of the neutrinos points to a flux that is mostly of extragalactic origin \citep{2019ApJ...886...12A}. 
A number of astrophysical source types have been suggested to be responsible for the observed signal. In particular blazars, long-known as efficient cosmic accelerators, have been proposed as likely sources of high-energy neutrinos (\mbox{Refs.
\citep{Mannheim1993,2015RPPh...78l6901A,TavecchioGhiselini2015}} and references therein).

In this early phase of multi-messenger astronomy, where the available instrumentation is limited in precision and sensitivity, unambiguous associations would greatly benefit from the detection of high-energy neutrinos together with enhanced activity in some parts of the electromagnetic spectrum of the astrophysical counterpart \citep{Aartsen2018,2021arXiv210714632S}. This condition, however requires that  hadronic-related electromagnetic flares 
are strong enough to outshine the non-thermal radiation generated via different mechanisms, e.g., from accelerated electrons radiating in magnetic fields in the so-called leptonic scenarios \citep{Ghisellini1985,Maraschi1992,Sikora1994}.
One such association in space and time was described in Ref. \cite{2016NatPh..12..807K}, which reported a large flare in the \gr\, and other parts of the spectrum from the blazar PKS\,1424$-$418 in correspondence with the detection of a $\sim$2 PeV IceCube neutrino. The positional uncertainty of this event (15.9$^{\circ}$, 50\% radius), however, is so large that the a posteriori probability of a chance coincidence was estimated to be about 5\%, too large for an unambiguous association.

So far only one astronomical 
 object has been associated with a high-energy astrophysical neutrino with a significance larger than $3 \sigma$. This is the blazar TXS\,0506+056, also known as 5BZB\,J0509+0541, located in the relatively small uncertainty region of the neutrino IceCube-170922A, detected when the source was undergoing a \gr\, flare. This result was further strengthened by the discovery of an excess of several lower energy neutrinos from the same direction in a subsequent analysis of IceCube archival data \citep{neutrino,Dissecting}. 
 This so-called ``neutrino flare'' occurred when TXS\,0506+056 was in a period of low \gr\, activity, suggesting that the relationship between astrophysical neutrinos and \gr\, emission is not straightforward. 

\section{Blazars}

Blazars are the most powerful sources of persistent non-thermal radiation in the Universe \citep{Urry1995}. They are a special and rare type of Active Galactic Nuclei (AGN, \citep{AGNReview}) with unique characteristics such as the emission of highly variable radiation over the entire electromagnetic spectrum. The most peculiar aspect about blazars is that this radiation originates within a relativistic jet that moves away from the central supermassive black hole and is oriented in the direction of the Earth. This is the very special condition that is responsible for the distinctive features of these sources such as superluminal motion and fast variability (see Refs. \citep{Blandford1978,Urry1995,2014A&ARv..22...73F}).

A few thousand blazars are known. Although these sources have been mostly discovered in the radio, X-ray or \gr\, bands, they are conventionally sub-classified depending on their optical spectrum as Flat-Spectrum Radio Quasars (FSRQs) and BL Lacertae objects (or BL Lacs), with FSRQs showing broad emission lines like radio-quiet Quasi-Stellar Objects (QSOs), and BL Lacs being featureless or at most displaying very weak emission lines  \citep{2014A&ARv..22...73F}.
All blazars emit across the entire electromagnetic spectrum with an energy distribution that, in \nufnu\, vs $\nu$ space, displays two broad humps, one attributed to synchrotron radiation that rises from radio frequencies and peaks between the far IR and the X-ray band, and the second, more energetic one due to inverse Compton or other mechanisms, that peaks in the low or high-energy \gr\, band \citep{GiommiPlanck}. See Ref. \cite{GiommiXRTspectra} for examples of different types of spectral energy distributions (SEDs) including large amounts of multi-frequency data. The peak energy and the relative intensity of the two humps span a wide range of values and have been used to classify blazars depending on the distribution of their power output.
On the basis of the rest-frame frequency of their low-energy hump (\nup) blazars are classified into low- (LBL/LSP: \nup~$<10^{14}$~Hz), intermediate- (IBL/ISP:
$10^{14}$~Hz$ ~<$ \nup~$< 10^{15}$~Hz), and high-energy
(HBL/HSP: \nup~$> 10^{15}$~Hz) peaked sources respectively
\citep{Padovani1995,Abdo_2010}. Extreme blazars are defined by \nup~$> 2.4
\times 10^{17}$~Hz (see Ref. \citep{biteau2020} for a recent review). In this paper we use the nomenclature LBL, IBL, and HBL as in Ref.  \citep{GiommiXRTspectra}, which is an extension of the codification originally defined in Ref. \cite{Padovani1995} 
applied to all blazar types, FSRQs and BL Lacs.

Although extremely rare among optical sources, blazars are by far the most common type of objects  detected by \gr\, telescopes outside the Galactic plane \citep{4FGL,4FGLDR2,2019A&A...627A..13B}. Given that neutrino production is inevitably associated with the generation of \grs, this peculiarity makes blazars natural candidate neutrino sources. From an observational perspective the situation is however complex since \grs\, can also be generated in purely leptonic contexts (e.g., synchrotron self-Compton or external inverse Compton \citep[][]{Maraschi1992,GhiselliniTavecchio2009}), and \grs\, associated to neutrino production could be completely or largely absorbed before they leave the emission region.


\section{Astrophysical Neutrinos and Blazars}


\textls[-15]{Blazars have long been considered as probable sources of  astrophysical neutrinos~\mbox{\cite{Stecker1991,Mannheim1992}}, but statistical searches for neutrino emitters have become possible only recently when sufficient data has been accumulated, mostly from IceCube observations. 
Investigations of this type have been conducted largely comparing neutrino arrival directions with lists of bright radio or \gr\, sources or with catalogues of well known blazars and other types of AGN. 
A search for point sources in the IceCube 10 year data collection, which includes events with energy typically \gsim 1 TeV, found an excess of neutrinos at the $2.9\,\sigma$ 
level from the direction of the local ($z = 0.004$) Seyfert 2 galaxy NGC\,1068 
and a $ 3.3\,\sigma$ excess in the northern sky due to significant p-values 
in the directions of NGC 1068 and three blazars: TXS\,0506+056, PKS\,1424+240, and 
GB6\,J1542+6129 \citep{2020PhRvL.124e1103A}. A similar work based on the database of the ANTARES neutrino observatory did not lead to the firm detection of point-like sources, with the most prominent excesses found in correspondence of the radio galaxy 3C 403 and the blazar MG3~J225517+2409 \citep{antares2019}.}

In the radio domain recent papers considered a number of data sets including a complete sample of 8~GHz sources selected from the very large baseline interferometry (VLBI)-based Radio Fundamental Catalogue (RFC), and the list of blazars monitored at various radio frequencies by the MOJAVE project and at the Owens Valley, the Metsähovi, and RATAN-600 Radio Observatories \cite{Plavin2020,Plavin2021,Hovatta2021,Desai2021}.
Possible correlations between radio sources and IceCube neutrinos have been reported in Ref. \cite{Plavin2020}, where it is concluded that four radio bright blazars, namely 3C 279, NRAO 530, PKS\,1741$-$4038, and 
PKS\,2145+067, all LBLs, 
are highly probable high-energy (E $> 200$ TeV) 
neutrino emitters. However 3C 279, the brightest and the only one of these blazars that is included in the list of objects used for the point source search in the 10-year IceCube data set, is not listed among the objects considered as likely neutrino emitters \citep{2020PhRvL.124e1103A}. In addition, a new analysis based on the same list of 3,411 bright radio sources and the updated sample of 10 year of IceCube data \cite{Zhou2021} did not confirm the results of Ref. \cite{Plavin2021}. 
A recent investigation aimed at the detection of flare neutrino emission in the 10 year IceCube data set \cite{Abbasi2021}
reported a time-dependent neutrino excess in the northern hemisphere at the level of 3$\sigma$ 
 associated with four sources: M87, a giant radio galaxy, TXS\,0506+056 and GB6 J1542+6129, two IBL blazars, and NGC 1068. 
 As noted in Ref. \cite{Giommidissecting} M87, which has \gr\, properties similar to those of HBL objects, is also the possible counterpart of the IceCube-141126A track-like event.

The most robust statistical evidence so far for an association between astrophysical neutrinos and blazars remains the case of TXS\,0506+056 \citep{Aartsen2018,neutrino,Dissecting}. This IBL/HBL type blazar is 
located well within the $\sim$ 0.5$^{\circ}$ (90\% containement) error region of the $\sim$0.3 PeV event IceCube-170922A that was detected in correspondence with enhanced \gr\, activity from the source. This already significant result was further strengthened by the detection of a 3.5$\sigma$ excess from the same direction found in a subsequent search of IceCube archival data \citep{neutrino,Dissecting}. The corresponding $13\pm5$ neutrinos were detected between September 2014 and March 2015, when TXS\,0506+056 was not undergoing strong \gr\, activity, implying that the relationship between astrophysical neutrinos and \gr\, emission is likely complex.
A similar, although less striking, event is the case of 3HSP J095507.9+35510, a blazar of the HBL type that is located in the error region of the highly energetic neutrino IceCube-200107A, that was detected during a strong X-ray flare \citep{2020A&A...640L...4G}.
 
\textls[-15]{To find similar cases in existing neutrino data and using all the available multi-frequency information Giommi et al. have carried out a systematic detailed study of 70~public IceCube high-energy neutrino tracks that are well reconstructed and away from the Galactic
plane~\citep{Giommidissecting}. This effort led to a $3.2\,\sigma$ (post-trial) correlation excess with $\gamma$-ray detected IBLs and HBLs, while no excess was found in correspondence of LBL blazars. 
This result, together with previous findings, consistently points to growing evidence for a connection between some IceCube neutrinos and IBL and HBL blazars. 
Moreover, several of the 47 IBLs and HBLs listed in Table 5 of Ref. \cite{Giommidissecting}, are expected to be new neutrino sources waiting to be identified. Further progress requires optical spectra, which are needed to measure the redshift, and hence the luminosity of the source, determine the
properties of the spectral lines, and possibly estimate the mass of the central black hole, $M_{\rm BH}$, which is the purpose of ``The spectra of IceCube Neutrino (SIN) candidate sources'' project. }

In this framework Paiano et al. \citep{Paiano2021} presented the spectroscopy of a large fraction of the objects 
selected in Ref. \cite{Giommidissecting}, which, together with results taken from the 
literature, covered $\sim 80$\% of that sample. This was the first publication 
of the SIN project 
whose aim is threefold: (1) to determine the nature of these sources; (2) to model their SEDs 
using all available multi-wavelength data and subsequently the expected neutrino
emission from each blazar; (3) to determine the likelihood of a connection 
between the neutrino and the blazar using a physical model for the 
blazar multi-messenger emissions, as done, for example, in Refs.
\cite{Petropoulou2015,Petropoulou2020}.
In the second SIN paper \cite{Padovani2021}, the sources studied in Ref. 
\cite{Paiano2021} have been characterised to determine their real nature, and also see if these sources 
are any different from  the rest of the blazar population. Of particular relevance 
here are the so-called ``masquerading'' BL
Lacs. Padovani et al. \citep{Padovani2019} showed, in fact, that TXS\,0506+056, the first plausible
non-stellar neutrino source is, despite appearances, {\it not} a blazar of
the BL Lac type but instead a masquerading BL Lac. This class was
introduced in Refs.~\cite{Giommi2012,Giommi2013} (see also Ref.
\cite{Ghisellini2011}) and includes sources which are in reality 
FSRQs 
whose emission lines are washed out by a very bright, Doppler-boosted jet
continuum, unlike ``real'' BL Lacs, which are {\it intrinsically}
weak-lined objects. This is relevant because ``real'' BL Lacs and FSRQs
belong to different physical classes, i.e. objects {\it without} and {\it
  with} high-excitation emission lines in their optical spectra, referred
to as low-excitation (LEGs) and high-excitation galaxies (HEGs) \cite{AGNReview}. 
TXS\,0506+056, being
a HEG, therefore benefits from several radiation fields external to the jet
(i.e. the accretion disc, photons reprocessed in the broad-line region or
from the dusty torus), which, by providing more targets for the protons
might enhance neutrino production as compared to LEGs. This makes
masquerading BL Lacs particularly attractive from the point of view of
high-energy neutrinos. Padovani et al. \cite{Padovani2021} have found that the sample considered in \cite{Giommidissecting} has a fraction $> 25\%$ and 
possibly as high as 80\% of masquerading sources, which is tantalizing.

\section{Blazars of Different Types}
\label{blazars}

In the previous section we have shown that experimental evidence is accumulating in favour of blazars being likely neutrino counterparts.
Motivated by the possibility that some specific blazar sub-classes may play a role in the emission of astrophysical neutrinos, in this section we use the best available samples and multi-frequency data to review similarities and differences among blazars sub-types. 

By definition, LBLs, IBLs and HBLs only differ by the value of  \nup\, in their SEDs. This spread could simply reflect the maximum energy at which particles are accelerated or could be due to other physical processes that determine the shape of the SED. 

\subsection{Blazar Samples}

At present the largest available lists of confirmed blazars are the following: 

\renewcommand{\theenumi}{\roman{enumi}}

\begin{enumerate}[leftmargin=2.2em,labelsep=0mm]
    \item The BZCAT catalogue, fifth edition \cite{Massaro2015}, which includes over 3,500 objects classified as BZQ (that is showing broad lines typical of QSOs), BZB (BL Lacs with no or very weak lines), BZU (blazars of uncertain type), and BZG (blazars where the optical/IR data in the SED reveals the presence of the host galaxy). BZCAT, being an heterogeneous list of blazars, is not a flux limited sample directly usable for statistical purposes;
    \item The 3HSP catalogue, which includes 2,013 HBL sources \citep{3HSP} with radio flux larger than $\sim$1 mJy. The sample is not complete at low radio flux densities values;
    \item The 4LAC-DR2 sample of \gr-selected AGN \citep{4LAC,4LAC-DR2}, which includes over 3,500 blazars of all types, many of which also included in the BZCAT and 3HSP catalogues.
\end{enumerate}

\subsubsection*{A Sample of IBL Blazars.}

\textls[-15]{No well-defined flux limited samples of IBL blazars currently exists. The only large list available is the subset of {\it Fermi} 4FGL-DR2 \gr\, sources classified as ISPs in the 4LAC-DR2 paper \citep{4LAC,4LAC-DR2}. The SED classification  of 4LAC-DR2 is however not always reliable, especially for IBL sources since at IR/optical frequencies, where IBLs peak, there are emission components not related to the jet, such as the host galaxy or the blue bump, that can sometimes lead to large inaccuracies in the estimation of \nup.
To assemble a large and accurately determined sample of \gr\,-detected IBL blazars we have built the SED of a large fraction of {\it Fermi} 4FGL-DR2 blazars through the Open Universe VOU-Blazar tool V1.94  \citep{VOU-Blazars}, which uses 71 multifrequency catalogues and spectral databases. In particular, we have considered the following samples: (a) all sources classified as ISPs in the 4LAC-DR2 catalogue; (b) blazars classified as HSPs in 4LAC-DR2 that are not included in the 3HSP sample \citep{3HSP}; (c) all sources classified as LSPs in 4FGL-DR2 
with \nup $>10^{13}$ Hz.
A careful visual inspection of the SED of each candidate allowed us to remove the components that can be related to non-jet emission. A fit to the average multi-frequency non-thermal components led to a sample including 482~sources with \nup~between $10^{14}$ and $10^{15}$ Hz, and radio flux densities ranging from a few mJy to over 6 Jy at 1.4 GHz. 
Particular care was also taken in the verification of the redshift of each object as in many cases this parameter was estimated by automatic software that was run on a large number of optical spectra. In some cases warning flags generated by the code have not been taken into account and wrong or inaccurate redshifts have been included in on-line sites and reported in the literature.
Our verification was done by visually inspecting published optical spectra, or online SDSS-DR16 spectra, when available. This
was useful and necessary since a number of sources with reported medium-high redshifts values in fact showed featureless optical spectra. }

Note that this sample of IBL blazars will evolve somewhat in the future depending on the availability of additional multi-frequency data.
The table including the details of each IBL blazar in the sample  will be made available via the Open Universe platform \mbox{(\url{https://openuniverse.asi.it}, accessed on 11 December 2021}).
By construction, this is a \gr\, flux-limited sample. However, considering the observed range of radio to \gr\, flux ratios for IBLs and the sensitivity of the 4FGL-DR2 catalogue, we expect it to include almost the totality of IBL blazars with radio flux densities $\gsim$ 150--200 mJy. 

\subsection{LBLs vs. IBLs vs. LBLs}

To investigate possible intrinsic differences between LBL, IBL, and HBL blazars we must use samples that are reasonably complete above a common flux limit in the same energy band.
Since all blazars in the available lists are radio sources with similar radio spectra, we use a conservative radio flux density limit of 200 mJy, which is sufficiently large to ensure that no objects above this limit are missed in the catalogues listed above. 

The IBL and HBL samples can be defined in a straightforward way, since they simply consists of the subset of sources with radio flux density $\ge 200$ mJy. 
For the sample of LBL sources we have taken all objects in the BZCAT that are above this flux density limit and are not included in the 3HSP catalogue or in our IBL sample.
To ensure uniformity in the radio data, and to avoid the complications of the Galactic plane, we consider the part of the sky covered by the NRAO VLA Sky Survey (NVSS: \citep{Condon1998}) (Dec $\ge -$40$^{\circ}$) with Galactic latitudes  |b| > 10$^{\circ}$, corresponding to an area of 28,400 square degrees of sky.
These selection criteria resulted in HBL, IBL and LBL subsamples  that include 38, 114, and 1370 sources, respectively.
The total blazar number density N($>$200 mJy) therefore is (38 + 114 + 1370)/28,400 = 0.054 deg$^{-2}$, a value that 
is close to that derived from the LogN-LogS of the DXRBS survey \citep{DXRBS} for all blazars types. We can therefore assume that these lists are complete at a level that is certainly sufficient for the purpose of finding relevant differences in the three sub-samples.

\begin{figure}[H]

\includegraphics[width=6.5cm]{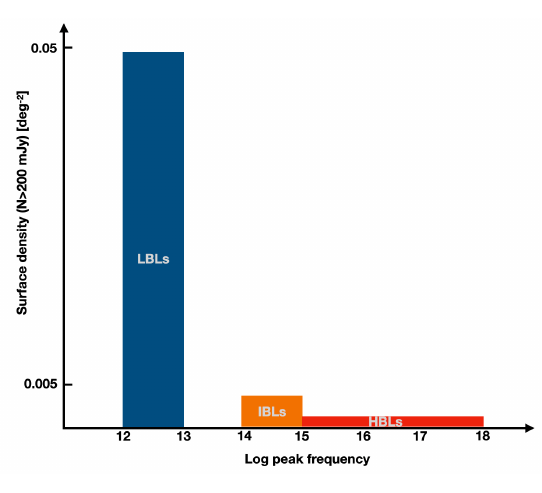}
\includegraphics[width=6.5cm]{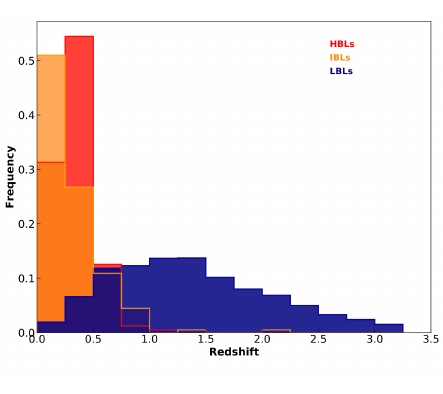}
\caption{The surface density (left side) 
 and redshift distributions (right side) of LBL, IBL and HBL blazars with radio flux density larger than 200 mJy.}
\label{fig:sp-z-distr}
\end{figure}

The relative space density of radio flux-limited ($f_{\rm radio} >$ 200 mJy) LBLs, IBLs and HBLs is shown on the left side of Figure \ref{fig:sp-z-distr}. A large gap between LBLs and other blazars types is clearly present. LBLs are over a factor 10 more abundant than IBLs, which are instead only about a factor of two more abundant than HBLs.
In this plot LBL sources are all confined in the Log(\nup) interval 12--13. Even though some blazars with Log(\nup) between 13 and 14 likely exist, from our experience objects possibly located in this frequency interval are rarely observed and their \nup\, is very difficult to estimate because 
the flux at these energies is often contaminated by components that are unrelated to the jet, like the dusty torus or emission from the host galaxy.
The low number of sources in this frequency bin marks a clear discontinuity between LBLs and the other types of blazars. 

\vspace{-3pt}

Another striking discontinuity is evident in the redshift distributions shown on the right side of Figure \ref{fig:sp-z-distr}, with the redshifts of both IBLs and HBLs being mostly confined to low values (z $\lsim 1.0$ and peaking below z = 0.5), while those of LBLs reach values larger than 3, with a mean value of 1.44 and a dispersion of 0.8. 
The large frequency of IBLs in the first redshift bin is due to the still relatively large number of these sources with no redshift estimation ($\sim30\%$) due to featureless or unavailable optical spectra, and which therefore are assigned the value of 0.
The number of IBL sources with zero redshift will most likely significantly decrease if photometric redshift estimations based on host galaxy signatures in the optical/IR parts of the SED, are carried out as in the case of the 3HSP sample~\citep{3HSP}. 
This sharp discrepancy in the redshift distributions reflects the well known different cosmological evolution between LBL (mostly FSRQs) and HBL blazars. 
The cosmological evolution of LBLs is strong and similar to that of optically or X-ray selected AGN, which are largely radio quiet QSOs \citep{DXRBS,Hasinger2005}. HBLs instead are known to show negative or no evolution \citep{DXRBS,Giommi1999,Rector2000}. No conclusive explanation for this low level of evolution in HBLs has been found yet. Since the redshift distribution of IBLs is similar to that of HBLs (Figure \ref{fig:sp-z-distr}, right side) it is reasonable to assume that the evolution of IBL blazars is similar or identical to that of HBLs.

\begin{figure}[H]
\includegraphics[width=6.6cm]{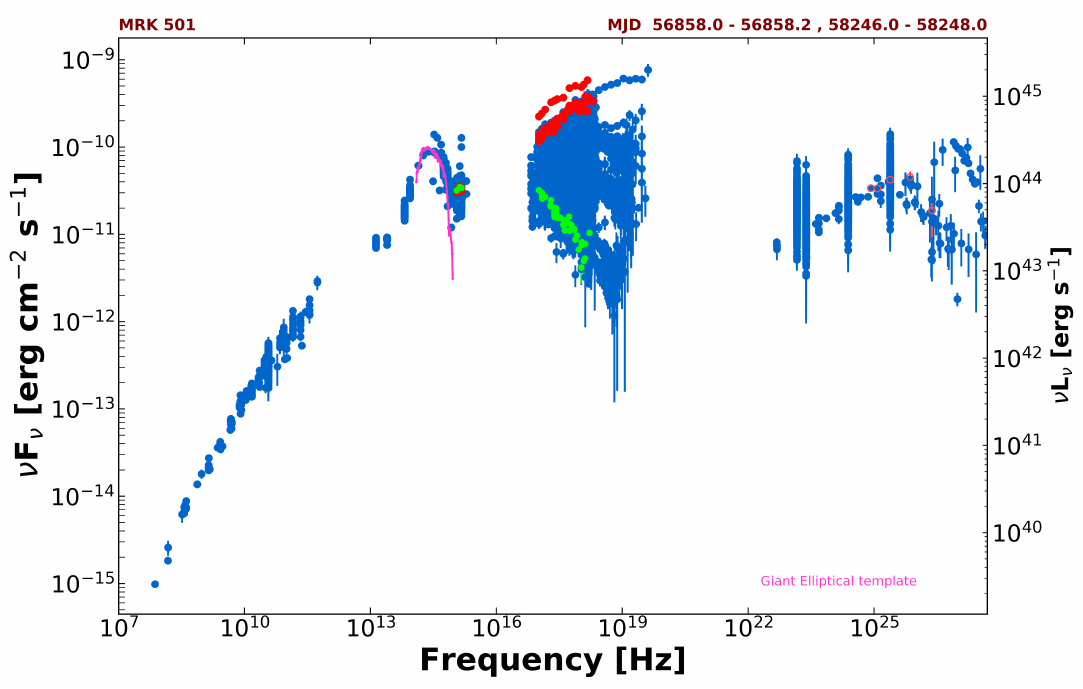}
\includegraphics[width=6.6cm]{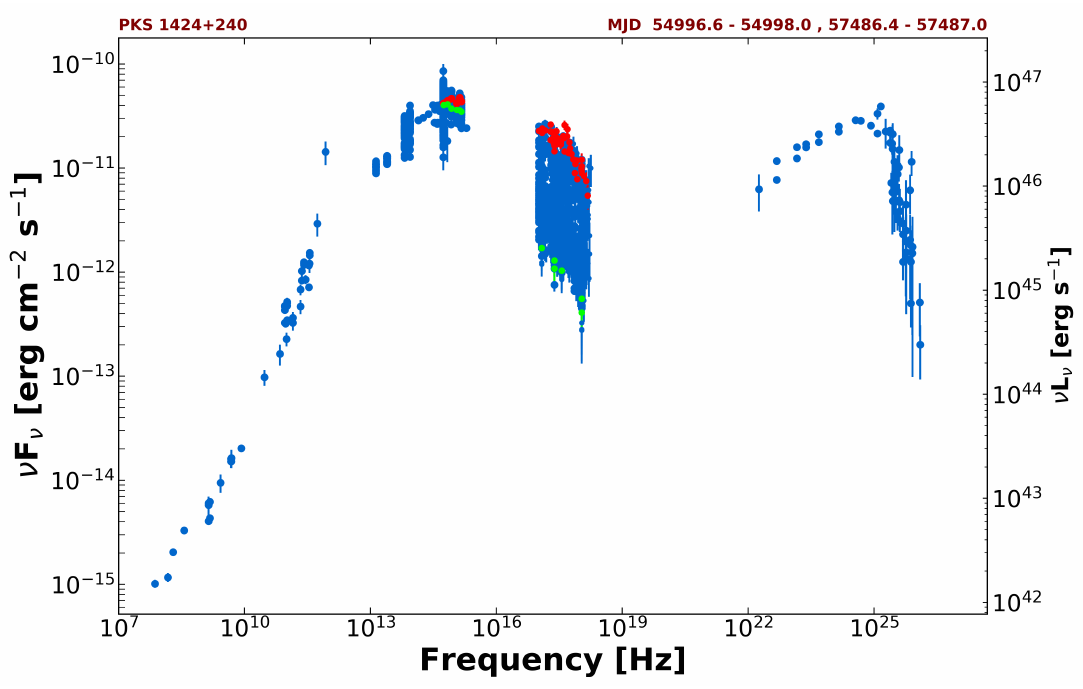}
\caption{The SED of the HBL 
 Mrk 501 (left) and the IBL PKS\,1424+240 (right) illustrating that most 
of the variability in these objects is in the X-ray band and translates into large shifts of 
the \nup~energy. Red and green colours represent Swift UVOT and XRT simultaneous data during high 
and low luminosity states respectively. 
\label{fig:iblhblseds}}
\end{figure}

\begin{figure}[H]
\includegraphics[width=6.6cm]{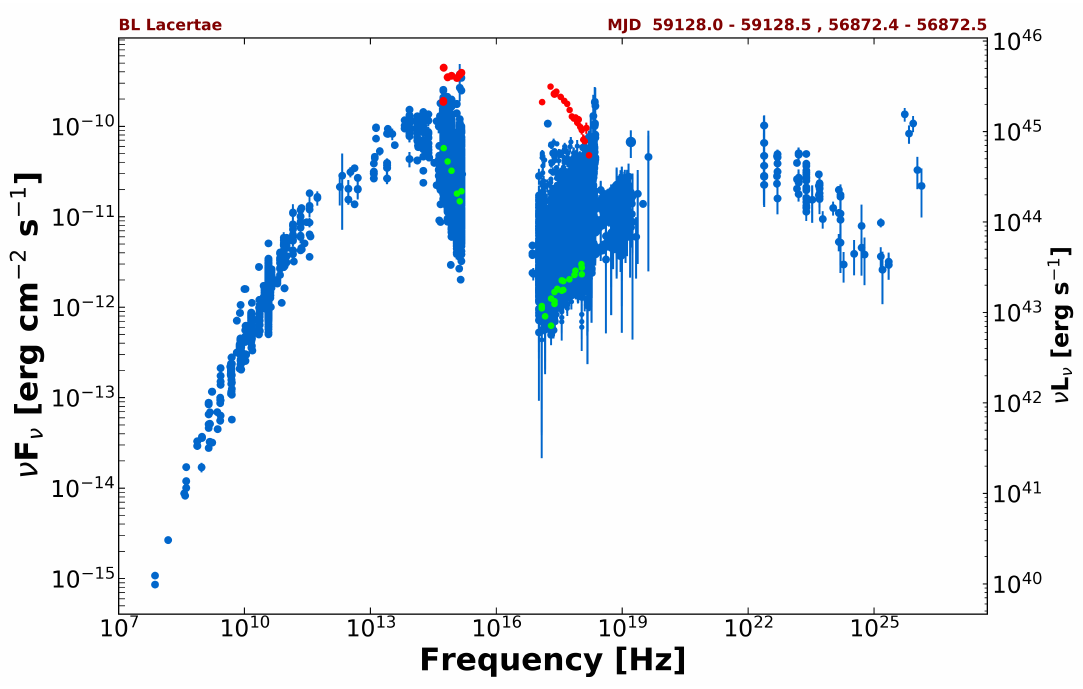}
\includegraphics[width=6.6cm]{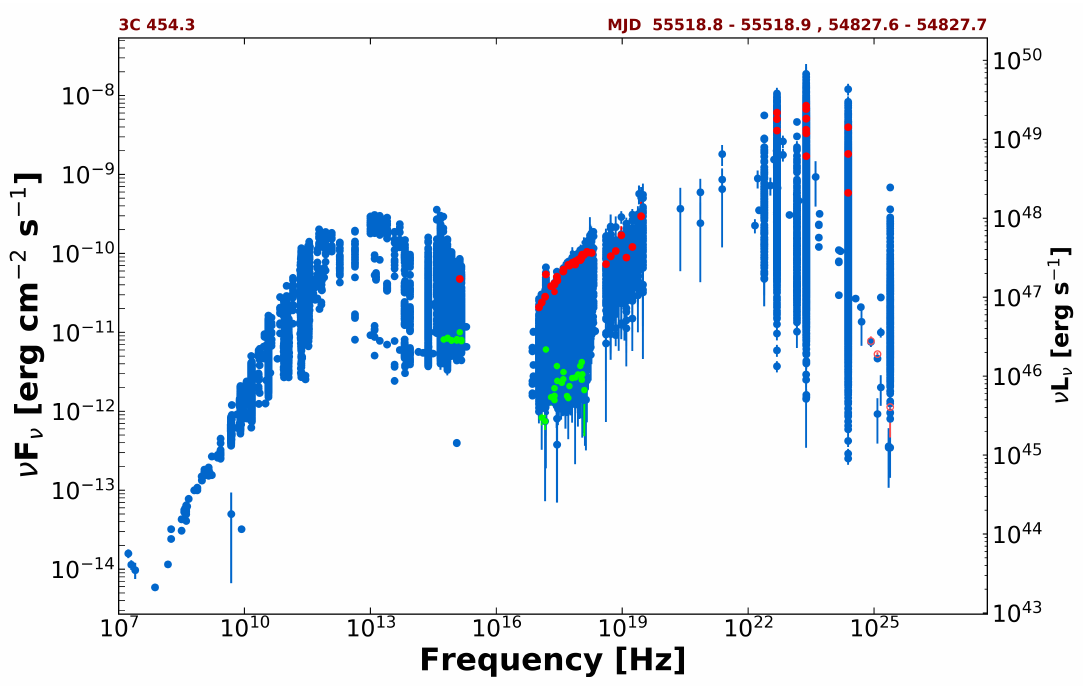}
\caption{The SED of the IBL source BL Lac (left), which shifted \nup\, by a large amount 
during a strong flare, and the LBL blazar 3C 454.3 (right), which shows large variability 
of similar amplitude from far IR to the X-ray band, and no large \nup\, shift. Red and 
green colours represent Swift UVOT and XRT simultaneous data during high and low luminosity 
states respectively.  
\label{fig:ibllblseds}}
\end{figure}

LBLs and IBLs/HBLs exhibit sharp differences also from the point of view of SED variability.
Large changes of \nup\, up to factors of 100 or more are commonly observed in HBL blazars (see Figure 4 of Ref. \cite{GiommiXRTspectra}) but not in LBLs. Most of the flux variability associated to \nup\, shifts in HBLs occurs in the X-ray band and is not accompanied by equivalent simultaneous changes at UV or at lower energies. Examples of this behaviour are illustrated in Figure \ref{fig:iblhblseds}
where the data corresponding to high and low states appear in red and green respectively, and all the available data are shown in blue. IBL blazars also show large \nup\, changes which cause them to occasionally become HBLs during high states. One such example is shown on the left side of Figure \ref{fig:ibllblseds} for the case of BL Lac, which shifted \nup\, from typical values around $10^{14}$ Hz to $\sim 10^{16}$ Hz during a large flare~\citep{BLLacPrince,BLLacSahakyan}. Several other IBL/HBL transitional sources are known; some of these, e.g., OJ 287, TXS\,0506+056, PKS\,2005$-$489, PG 1553+113, are among the blazars that have been frequently observed by Swift, and their SEDs can be found in Ref. \cite{GiommiXRTspectra} and online at \url{https://openuniverse.asi.it/blazars/swift/} (accessed on 11 December 2021).

\vspace{-6pt}

In LBL blazars \nup\, remains fairly stable between $\sim 10^{12}$ and $\sim 10^{13}$ Hz even during large flares since flux variability in these sources seems to occur almost a-chromatically, with similar amplitude at all frequencies from far IR to UV and X-rays \citep{GiommiXRTspectra,GiommiPlanck}. One example of such behaviour is shown in the right side of Figure \ref{fig:ibllblseds} for the case of 3C 454.3. Other examples of LBL blazars with well populated SEDs that follow the same behaviour are 3C 279, PKS\,0235+164 and CTA 102  \citep{GiommiXRTspectra} ({\url{https://openuniverse.asi.it/blazars/swift/}}, \mbox{accessed on 11 December 2021}).
Blazars also show differences in jet kinematics. LBLs exhibit a wide distribution 
of apparent jet speeds, with values reaching 30--40c, with jet speeds instead being typically
$\lesssim$5c and $\lesssim$10c for HBL and IBLs respectively \citep{Lister2019}. 
An upcoming paper (Padovani et al., 2022, in preparation) concludes, based on VLBI data,  
that the two neutrino candidates TXS\,0506+056 and PKS\,1424+240 have overall parsec-scale 
properties similar to HBLs and different from those of LBLs. 

LBL and HBL blazars also differentiate in their optical polarisation properties, as demonstrated in Ref. \cite{Jannuzi1994} where the authors studied a sample of
X-ray selected BL Lacs (mostly HBLs) and concluded that these objects show a lower level of optical polarisation (with a maximum of $\sim$10\%) and lower duty cycle than radio selected BL Lacs (that are mostly LBLs where the optical flux is dominated by the jet and not by the blue bump and broad lines emission), which are instead characterised by polarisation levels of 30--40\% and a high duty cycle \citep{ImpeyTapia1990}.

Based on the empirical evidence described above, which reveals strong similarities between IBLs and HBLs properties, and large differences from LBLs, we suggest that blazars come only in two main flavours: LBLs, and IBLs plus HBLs combined, which we propose to collectively call IHBLs. 
For practical purposes, blazars can be assigned to one of the two sub-classes on the basis of a single \nup\, value, with LBLs being those with \nup~$< 10^{13.5}$ Hz and IHBLs those with \nup~$> 10^{13.5}$~Hz.

We stress that LBLs and IHBLs are not simply sub-classes of blazars bureaucratically defined by the value of an observational parameter, but  reflect deeply different intrinsic properties, such as cosmological evolution, demographic characteristics, broad-band spectral variability, optical polarisation, and possibly the physical conditions in the emitting regions that in IHBLs might host efficient proton acceleration.
Since the cosmological evolution of LBLs  is similar to that of optically and X-ray selected radio quiet QSOs, the presence of LBL relativistic jets, at least in first approximation, occurs in QSOs independently of cosmic epoch.
The same does not happen in IHBL jets, which instead follow a different evolution. If proton acceleration occurs only (or mostly) in IHBL source it could be that the conditions that enable proton acceleration drive the evolution in these sources.

\subsection{Transient Blazars and Neutrino Astronomy}

The \gr\, source 4FGL\,J1544.3$-$0649 is a remarkable blazar that remained unnoticed at high energies until May 2017 when it showed a transient-like behaviour brightening to such a level to be detected by {\it Fermi}-LAT and the MAXI X-ray sky monitor. The source remained bright for a few months exhibiting an SED typical of bright HBL blazars \cite{TransientBlazar}.
This discovery suggests the existence of a population of still undiscovered objects that can occasionally flare and become strong X-ray and \gr\, sources.  
If X-ray flares are indeed associated to proton acceleration and neutrino emission, as suggested in Ref. \cite{2021ApJ...906..131M}, these sources could play a significant role in neutrino astronomy. Considering that the expected neutrino flux from 4FGL\,J1544.3$-$0649 is the fifth largest in the list of 66 bright blazars considered in Ref. \cite{2021arXiv210714632S}, blazars that are currently uncatalogued and below detectability in the X-ray and \gr\, bands could even be a major population of neutrino sources. In this case the identification of the counterpart of many neutrino events would be a very difficult task.

The importance of this possible contribution clearly depends on the abundance of this population of blazars, how often large transient events occur, and how long sources remain in a bright phase.
A reliable estimate of the space density and duty cycle of these elusive objects would only be possible through all-sky X-ray or \gr\, monitors that are significantly more sensitive that those currently in operation.
Sensitive all-sky monitors would also easily discover possible high-energy flares from objects in the large localisation errors of neutrino events, providing ``smoking-gun'' evidence for neutrino-blazar association. 

\section{Summary and Discussion}

Neutrino astronomy is still in a nascent phase, a status characterised by consolidated results on the detection of neutrinos of astrophysical origin, but also by the lack of indisputable evidence about the nature of their electromagnetic counterparts. Some results in this field are sound, such as the absence of anisotropy in the arrival directions of high-energy IceCube neutrinos, which implies a dominant population of extragalactic sources, although a minor Galactic component cannot be excluded. 
Considering specific associations with known cosmic electromagnetic sources we have shown that a growing number of papers are reporting possible associations of astrophysical neutrinos with AGN, particularly blazars of different types. Substantial theoretical work also demonstrated that neutrino emission mechanisms consistent with the existing data can be accommodated in physical situations that are typical of blazars \cite{Mannheim1993,2015RPPh...78l6901A,TavecchioGhiselini2015,Petropoulou2015,Petropoulou2020,Rodrigues2021}.

In the following we make some considerations on the nature of the proposed associations with significance larger than 2 $\sigma$ for different types of blazars. 

\begin{figure}[H]
\includegraphics[width=10.0cm]{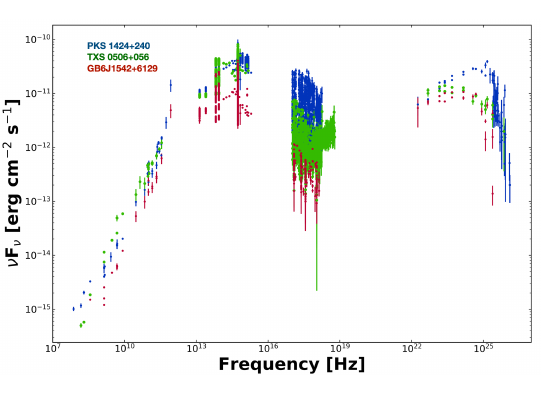}
\caption{The SED of the 
 three blazars reported as possible neutrino sources in \citep{2020PhRvL.124e1103A}. They are all of the IBL type with remarkably similar flux and overall shape, except in the X-ray band, where small differences in \nup\,, large variability and non-simultaneous observations cause a large scatter. 
\label{fig:sedcomp}}
\end{figure}

As reported in the previous sections correlations of radio-bright AGN with astrophysical neutrinos have been sought extensively \citep{2016NatPh..12..807K,Plavin2020,Plavin2021,Hovatta2021,Desai2021}.
Recently, possible statistical associations with specific LBL sources (3C 279, NRAO 530, PKS\,1741$-$038, and PKS\,2145+067) have been reported in Ref. \cite{Plavin2020}. These results however were not confirmed by similar works~\citep{Zhou2021,Desai2021}.
Samples of bright radio sources tends to select LBL objects since IHBL blazars are largely underrepresented in these data sets. For example only 103 out of the 3411 sources ($\sim 3\%$) of the complete RFC sub sample with f$_{\rm r,8 GHz} > 150 $ mJy  used in Refs.~\cite{Plavin2021,Zhou2021} are IHBL blazars. A similar proportion is also evident in the surface density comparison plotted on the left part of Figure \ref{fig:sp-z-distr}, which refers to blazars with radio flux density larger than 200 mJy at 1.4 GHz.
Despite the large surface density of LBL blazars and the multiple statistical searches based on radio-bright sources, no stable  significant association with sources of this type has been found. 
The matching of astrophysical neutrino samples with the position of known \gr-detected blazars, regardless of their radio flux, resulted instead in frequent possible associations with blazars of the IHBL type. Numerous examples of this increasingly reported outcome exist. The following is a list of the most important cases.
\begin{itemize}
    \item A detailed study of the blazars located in the error regions of a sample of 70 well-reconstructed IceCube tracks found 47 IHBL blazars compared to a background expectation of 26.8, an overabundance equivalent to a 3.2 $\sigma$ post trial significance. No excess of LBL blazars was found \citep{Giommidissecting}. This is the statistically most compelling result in favour of IHBLs. Very recently Savard et al. \cite{Savard2021}, extending this work to 10 additional and newly detected neutrino tracks, confirmed that blazars from the 3HSP catalogue are  significantly overrepresented in neutrino error regions.
    
    \item The famous blazar TXS\,0506+056, currently considered the most reliable source of IceCube neutrinos \citep{Aartsen2018,neutrino,Dissecting}, is an IHBL source.
    \item A chance probability of $\lsim$1 per cent was found in Ref. \cite{Padovani2016} where the authors compared the positions of  extreme (IHBL) blazars with those of a sample of IceCube neutrinos. Other types of blazars gave null results. 
    \item The search for point-like sources in the 10 year IceCube sample \citep{2020PhRvL.124e1103A} resulted in a 3.3\,$\sigma$ excess over the expected background associated with the bright Seyfert 2 galaxy NGC 1068 and the three blazars PKS\,1424+240, TXS\,0506+056, and GB6~J1542+6129, all of which are bright (f$_{\rm radio}$ \gsim 200 mJy) IHBL objects with very similar overall SEDs, as illustrated in Figure \ref{fig:sedcomp}. A study aimed at finding additional similarities among these three sources will be presented by Padovani et al. (2022), in preparation.
    \item The IHBL blazar 3HSP\,J095507.9+35510, located in the error region IceCube-200107A, 
    was undergoing a strong X-ray flare at the time of the neutrino arrival \citep{2020A&A...640L...4G}. 
    \item The blazar MG3 J225517+2409, a bright IHBL source, has been proposed as the possible counterpart of 5 neutrinos detected by the ANTARES neutrino observatory \citep{antares2019} and the track event IceCube-100608A \citep{Giommidissecting}. 
    \item The 2016 detection of three neutrinos within 100 s of each other from the same direction~\citep{threeneutrinos} was possibly associated with 3HSP\,J013632.5+390559, a bright IHBL blazar  \citep{Giommidissecting}.
    \item A very recent search for neutrino flare emission in the IceCube 10 year data set~\citep{Abbasi2021} reported possible flares at the level of 3$\sigma$ from NGC\,1068, two IHBL blazars TXS\,0506+056, GB6\,J1542+6129, and M87, a giant radio galaxy with \gr\, properties similar to that of IHBL objects that was already noticed as a possible counterpart of the track IceCube-141126A \citep{Giommidissecting}. The IHBL 1ES1959+650, the next object in order of significance in the list of probable flaring sources of Ref. \citep{Abbasi2021}, was also noticed as a possible counterpart of three neutrinos detected by AMANDA, the predecessor of IceCube, in 2002 in correspondence of a rare high-energy \gr\, flare \citep{Halzen2021}. 
\end{itemize}
\vspace{3pt}

If the several hints linking IHBLs to astrophysical neutrinos genuinely reflect a particular physical situation in these sources, it could be that the characteristics that distinguish them (e.g., high \nup, no cosmological evolution, slow jet, strongly chromatic variability, low level of optical polarisation) may simply be the observational consequences of proton acceleration to very high energies. 
The highly variable UV or X-ray radiation that defines IHBLs could then be the direct result of proton synchrotron emission as proposed in Refs.~\cite{2021ApJ...906..131M,2021arXiv210714632S} or of radiation produced by the secondary particles generated in photo hadronic interactions that lose energy on timescales that depend on the specific physical conditions of the emitting region or of its evolution after flares. In this view the existence of IHBL blazars would be strictly connected to proton acceleration in AGN. The still poorly understood low level of cosmological evolution of IHBLs compared to other blazars, might also be connected with the conditions that lead to proton acceleration. 

\vspace{-10pt}

An origin of high-energy neutrinos from a non-evolving and low-density population of sources, such as IHBLs, easily meets the constraint for UHECRs sources not to overproduce the observed \gr\, background \citep{2017AIPC.1792f0005L}.
Intermittent proton acceleration occurring in persistent IHBLs, or in transient blazars like 4FGL\,J1544.3-
0649, would further ease the constraint on the \gr\, background.

Neutrino multi-messenger astronomy is an exciting new field in rapid evolution where new 
observational data, possibly of improved localisation accuracy, need to be accumulated 
to confirm or disprove the hypothesis that IHBL blazars are the preferential site of 
proton acceleration. Forthcoming observatories like 
KM3NeT ({\url{https://www.km3net.org/}}, accessed on 11 December 2021) and 
Baikal-GVD ({\url{https://baikalgvd.jinr.ru/}}, accessed on 11 December 2021) under water neutrino telescopes, the
Pacific Ocean Neutrino Experiment (P-ONE) \cite{PONE}, and the future more sensitive 
IceCube-Gen2 ({\url{https://icecube.wisc.edu/science/beyond/}}, accessed on 11 December 2021)
instrument, ideally operating in parallel with a new generation of sensitive all-sky X-ray and \gr\, detectors, are the facilities that will bring neutrino astronomy into the next 
phase, settling the question about the role of the different types of AGN in terms of their neutrino and UHECR emission.

\vspace{6pt} 



\authorcontributions{Both authors contributed equally to this work.}

\funding{This research received no external funding.}

\institutionalreview{Not applicable.}

\informedconsent{Not applicable.}


\dataavailability{\textls[-15]{The multi-frequency data of all SEDs presented in this work are available through the tools and web pages of the Open Universe platform, e.g., the VOU-BLazars application \citep{VOU-Blazars} and the interactive table at \url{https://openuniverse.asi.it/blazars/swift/} (accessed on 11 December 2021).}}

\acknowledgments{We thank Bia Boccardi for a careful reading of the paper. We acknowledge the use of data, analysis tools and services from the Open Universe platform, the National Extra-galactic Database (NED), and the bibliographic services of the Astrophysics Data System (ADS).}

\conflictsofinterest{The authors declare no conflict of interest.} 

\end{paracol}

\reftitle{References}

\end{document}